\documentclass{PoS}
\usepackage{amsmath}
\usepackage{amssymb, fontenc, times, mathptmx, graphicx}
\usepackage[utf8]{inputenc}

\title{Simultaneous explanation of $K$ and $B$ anomalies in vectorlike compositeness}

\ShortTitle{Explaining $K$ and $B$ anomalies in vectorlike compositeness}

\author{Shinya Matsuzaki\\
        Center for Theoretical Physics and College of Physics, Jilin University, Changchun, 130012, China.\\
        Department of Physics, Nagoya University, Nagoya 464-8602, Japan.\\
        E-mail: \email{synya@jlu.edu.cn}}

\author{\speaker{Kenji Nishiwaki}\\
        School of Physics, Korea Institute for Advanced Study~(KIAS), Seoul 02455, Republic of Korea.\\
        Ru{\dj}er Bo{\v{s}}kovi{\'c} Institute, Division of Theoretical Physics, Bijenicka cesta 54, 10000 Zagreb, Croatia.\\
        E-mail: \email{knishiw@irb.hr}}

\author{Kei Yamamoto\\
        Graduate School of Science, Hiroshima University, Higashi-Hiroshima 739-8526, Japan.\\
        Physik-Institut, Universit\"at  Z\"urich, CH-8057 Z\"urich, Switzerland.\\
        E-mail: \email{keiy@hiroshima-u.ac.jp}}

\abstract{
We address the presently reported significant flavor anomalies 
in the $K$ and $B$ meson systems such as the CP violating Kaon
decay ($\epsilon'/\epsilon$) and 
lepton-flavor universality violation in
$B$ meson decays ($R_{K^{(*)}},$ and also commenting ${R_{D^{(*)}}}$),  
by proposing flavorful and 
chiral vector bosons as the new physics constitution at $\sim 1\,\mathrm{TeV}$.
Interestingly, if the new (composite) vector bosons are quite heavier than $\sim 1\,\mathrm{TeV}$,
we face a difficulty in addressing the anomaly in $\epsilon'/\epsilon$
consistently with the constraint from the $K^0$-$\overline{K^0}$ mixing.
Both of the anomalies can be addressed within $1\sigma$ confidence levels individually,
where the relevant parameter space will be investigated by the NA62 and KOTO experiments,
in addition to direct searches at the large hadron collider. 
}

\FullConference{Corfu Summer Institute 2018 "School and Workshops on Elementary Particle Physics and Gravity"\\
		(CORFU2018)\\
		31 August - 28 September, 2018\\
		Corfu, Greece}

\begin{document}

\section{Introduction}

The experimental anomalies in $B$ meson decays whose partonic processes are described
as $b \to s \mu^+ \mu^-$ and $b \to c \tau \overline{\nu}$,
which include the QCD-safe observables $R_{K^{(*)}} = \mathcal B({B} \to K^{(*)} \mu^+ \mu^-) / \mathcal B({B} \to K^{(*)} e^+ e^-)$
and $R_{D^{(*)}} = \mathcal B(\bar{B} \to D^{(*)}\tau\bar\nu) / \mathcal B(\bar{B} \to D^{(*)}\ell\bar\nu)$,
have been focused extensively and especially for the last several years.
See {\it e.g.,} Refs.~\cite{Descotes-Genon:2013vna,Descotes-Genon:2013wba,Altmannshofer:2013foa,Hiller:2014yaa,Altmannshofer:2014rta}
and Refs.~\cite{Datta:2012qk,Celis:2012dk,Crivellin:2013wna,Dorsner:2013tla,Sakaki:2013bfa} for earlier works, respectively.
We find works in similar contexts
(compositeness or warped extra dimension)~\cite{Biancofiore:2014wpa,Niehoff:2015bfa,Niehoff:2015iaa,Carmona:2015ena,
Barbieri:2015yvd,Megias:2016bde,Barbieri:2016las,Megias:2017ove,Buttazzo:2017ixm,Megias:2017vdg,
Cline:2017aed,Bordone:2017bld,DAmbrosio:2017wis,Blanke:2018sro,Bordone:2018nbg}
(see also discussions about loop corrections of vector leptoquarks~\cite{Crivellin:2018yvo,Aebischer:2018acj}).
It is noted that recently another measurement of the $b \to c \tau \overline{\nu}$ transition, namely the ratio
$R_{J/\psi} = \mathcal B(B_c \to J/\psi \tau\bar\nu) / \mathcal B(B_c \to J/\psi \mu\bar\nu)$ has beem measured and discussed
(see e.g.,~\cite{Watanabe:2017mip,Aaij:2017tyk}).

On the other hand in the Kaon sector, the observable, $\epsilon'/\epsilon$, which measures the direct CP violation in $K \to \pi\pi$ decays
has been focused since sizable discrepancy from the experimental data
[$\left(\epsilon'/\epsilon \right)_{\mathrm{exp}} = (16.6 \pm 2.3) \times 10^{-4}$]~\cite{Batley:2002gn,AlaviHarati:2002ye,Abouzaid:2010ny,Tanabashi:2018oca}
was reported in the standard model~(SM) as
$\left(\epsilon'/\epsilon \right)_{\rm SM} =  (1.06 \pm 5.07) \times 10^{-4}$~\cite{Kitahara:2016nld} (see also~\cite{Buras:2015yba})
adopting the first lattice calculation result reported by RBC-UKQCD collaboration \cite{Bai:2015nea}.
We find various studies on this subject~\cite{Blanke:2015wba,Buras:2015xba,Buras:2015kwd,Buras:2016fys, Tanimoto:2016yfy,Kitahara:2016otd, 
Endo:2016aws,Bobeth:2016llm,Endo:2016tnu,Crivellin:2017gks,Chobanova:2017rkj,
Endo:2017ums,Bobeth:2017ecx,Gisbert:2017vvj,Haba:2018byj,Chen:2018ytc,Chen:2018vog,
Buras:2018wmb,Matsuzaki:2018jui,Haba:2018rzf,Aebischer:2018rrz,Aebischer:2018quc,
Aebischer:2018csl,Chen:2018dfc,Chen:2018stt,Chen:2018lze,Buras:2018ozh,Marzo:2019ldg}.
This anomaly would suggest us another hint for surveying possibilities of physics beyond the SM through flavor physics.

We consider the possible scenario where the $B$ and $K$ anomalies are addressed simultaneously discussed in~\cite{Matsuzaki:2018jui}.
Here, the chiral-flavorful vectors (CFVs) are introduced as a 63-plet of 
the global $SU(8)$ symmetry, identified as 
the one-family symmetry for left-handed quarks and leptons 
in the SM forming the 8-dimensional vector, where the 63-plet is decomposed into
massive gluon-like, vector-leptoquark-like, $W'$- and $Z'$-like vector particles.
CFVs contribute to both kinds of the anomalies in the $B$ and $K$ meson sectors, where
we found the parameter space where both kinds of the anomalies can be addressed within $1\sigma$ confidence levels~(C.L.s) individually.

\section{Chiral-flavorful vectors}

\subsection{General aspects}

The CFVs' (denoted as $\rho$) couplings to the left-handed fermions in the SM are constructed in 
the one-family global-$SU(8)$ symmetric way as 
\begin{align} 
{\cal L}_{\rho f_Lf_L} 
&
{
= \sum_{i,j = 1}^{3} 
g_{\rho L}^{ij} \overline{f}_L^i \gamma^\mu 
\rho_\mu f_L^j 
}
\,, \label{rhoLff}
\end{align}
where $g_{\rho L}^{ij}$ denotes the (hermitian) couplings with the generation indices ($i,j$) {in the gauge eigenbases},  
and 
$f_L^i$ includes the {left-handed SM doublet quarks} ($q^{ic}_L=(u^{ic},d^{ic})_L^T$ 
with the QCD color index $c=r,g,b$) 
and {(left-handed)} lepton doublets ($l^i_L=(\nu^i, e^i)_L^T$) for the $i$th generation, which forms 
the 8-dimensional vector (the fundamental representation of the $SU(8)$) 
like
$f_L^i = (q^{i r}, q^{i g},q^{i b}, l^i)_L^T$.  
To manifestly keep the SM gauge invariance in the coupling form of Eq.(\ref{rhoLff}) leads to
the gauging of the global $SU(8)$ symmetry,
\begin{align} 
 D_\mu \rho_\nu = \partial_\mu \rho_\nu - i [{\cal V}_\mu, \rho_\nu] 
 \,, \label{D}
\end{align}
where the SM gauge fields $(G_\mu, W_\mu, B_\mu)$ for the 
$SU(3)_c \times SU(2)_W \times U(1)_Y$ symmetry 
are embedded in the $8 \times 8$ matrix form of 
${\cal V}_\mu$ as 
\begin{align} 
{\cal V}_\mu = 
\left( 
\begin{array}{c|c} 
{{\bf 1}_{2 \times 2} \otimes g_s G_\mu^a \frac{\lambda^a}{2}}
+ {\left( g_W  W_\mu \tau^{\alpha} 
+  \frac{1}{6} g_Y B_\mu \right) \otimes {\bf 1}_{3\times 3}}
& {{\bf 0}_{6 \times 2}} \\ 
\hline 
{{\bf 0}_{2 \times 6}} 
& 
g_W W_\mu^{{\alpha}} \tau^{{\alpha}} 
- \frac{1}{2} g_Y B_\mu \cdot {\bf 1}_{2 \times 2} 
\end{array}
\right) 
\,, \label{Vcal}
\end{align}
where $\lambda^a$ and $\tau^\alpha \equiv \sigma^\alpha/2$ $(\alpha=1,2,3)$ 
are Gell-Mann and (normalized) Pauli matrices, and 
$g_s, g_W$ and $g_Y$ the corresponding gauge couplings.  
It is convenient to classify the CFVs ($\rho$) in the $SU(8)$ adjoint representation 
by the QCD charges as 
\begin{equation} 
{\rho = 
\left( 
\begin{array}{cc} 
(\rho_{QQ})_{6\times 6} & {(\rho_{QL})_{6 \times 2}} \\ 
{(\rho_{LQ})_{2 \times 6}} & (\rho_{LL})_{2 \times 2} 
\end{array}
\right)}
\,,    
\end{equation}      
where $\rho_{QQ}, \rho_{QL}(=\rho_{LQ}^\dag)$, and $\rho_{LL}$ include 
color-octet $\rho_{(8)}$ (of ``massive gluon $G'$ type''), 
-triplet $\rho_{(3)}$ (of ``vector-leptoquark type''), 
and -singlet $\rho_{(1)^{(\prime)}}$ (of ``$W'$ and/or $Z'$ type''), which can further be 
classified by the weak isospin charges ($\pm, 3$ for triplet and $0$ for singlet).  
Thus, decomposing the CFVs with respect to the SM charges, we find 
\begin{align}
 \rho_{QQ} = 
 &
 \left[ \sqrt 2 \rho_{(8)a}^{{\alpha}} \left( {\tau^\alpha} \otimes {\lambda^{{a}} \over 2} \right) 
 + {1 \over \sqrt 2} \rho_{(8)a}^0 \left( {\bf 1}_{2\times2} \otimes {\lambda^{{a}} \over 2} \right) \right] \notag \\[0.5em]
 & 
 + \left[ {1 \over 2} \rho_{(1)}^{{\alpha}} \left( {\tau^\alpha} \otimes {\bf 1}_{3\times3} \right) 
 + {1 \over 2\sqrt 3} \rho_{(1)'}^{{\alpha}} \left( {\tau^\alpha} \otimes {\bf 1}_{3\times3} \right) 
 + {1\over 4 \sqrt 3} \rho_{(1)'}^0 \Big( {\bf 1}_{2\times2} \otimes {\bf 1}_{3\times3} \Big) \right] \,, \notag \\[1em] 
 \rho_{LL} = 
 &
 {1 \over 2} \rho_{(1)}^{{\alpha}} \left( {\tau^\alpha} \right) - {\sqrt 3 \over 2} \rho_{(1)'}^{{\alpha}} \left( {\tau^\alpha} \right) - {\sqrt 3 \over 4} \rho_{(1)'}^0 \Big( {\bf 1}_{2\times2} \Big) \,, \notag \\[1em]
 \rho_{QL} = 
 &
 \rho_{(3){c}}^{{\alpha}} \left( {\tau^\alpha} \otimes {\bf e}_c \right) + {1 \over 2} \rho_{(3) {c}}^0 \Big( {\bf 1}_{2\times2} \otimes {\bf e}_c \Big) \,, \notag \\[1em]
 {\rho_{LQ}} = & \Big( {\rho_{QL}} \Big)^\dag \,, 
 \label{rho:assignment}  
\end{align} 
where ${\bf e}_c$ denotes the 3-dimensional eigenvector in the color space.

Parts of CFVs are mixing with SM gauge bosons through
the mass mixing form, 
\begin{align} 
- \frac{2 m_\rho^2}{g_\rho} {\rm tr}[{\cal V}_\mu \rho^\mu],  
\label{mass-mixing} 
\end{align}
where $m_\rho$ is the mass scale for CFVs, and $g_\rho$ governs
the magnitude of the mixings (where $g_\rho$ is the corresponding gauge coupling
of the (partial) gauging of the $SU(8)$ global symmetry based on the hidden local symmetry formulation).
This term generates the flavor-universal couplings 
for the CFVs to {both of the left-handed and right-handed quarks/leptons}, 
where the magnitudes of induced interactions are evaluated as $\sim (g_{s,W,Y}^2/g_\rho)$.  
Due to the constraint from electroweak precision measurements, 
$g_\rho$ should be greater than $\mathcal{O}(1)$, where $g_\rho \sim {10} $ is a safe choice.
It is mentioned that not only the flavorful interactions in Eq.(\ref{rhoLff}), but also
the flavor universal interactions through the ${\cal V}$-$\rho$ mixing sizably contribute to
phenomena, especially for $\epsilon'/\epsilon$ and the muonic resonance production at the large hadron collider~(LHC),
in spite of the suppression factor $\sim (g_{s,W,Y}/g_\rho)$ for $g_\rho \sim 10$.
In general, this mixing effect generates mass splitting among the CFVs.
Nevertheless, now the magnitude is very small as $(g_{s,W,Y}/g_\rho) \ll 1$,
it is reasonable to treat all of CFVs being degenerated, $M_{\mathrm{CFVs}} \simeq m_\rho$.
See~\cite{Matsuzaki:2017bpp} for an ultraviolate-completed realization by a vectorlike confining gauge theory.

\subsection{The flavor-texture Ansatz}

For our purpose of addressing the anomalies in the $B$ and $K$ sectors,
we decided to introduce the flavored texture for the $g_{\rho L}^{ij}$ in Eq.(\ref{rhoLff}). 
The concrete form of the flavor texture is
 \begin{equation} 
 g_{\rho L}^{ij} 
 = 
  {\begin{pmatrix} 
  0 & g_{ \rho L}^{12} & 0 \\ 
  (g_{\rho L}^{12})^* & 0 & 0 \\ 
  0 & 0 & g_{\rho L}^{33}  
  \end{pmatrix}^{ij} \,, }
\label{grhoL}
\end{equation} 
in which the hermiticity in the Lagrangian of Eq.(\ref{rhoLff})  
has been taken into account (i.e. $g_{\rho L}^{21} = (g_{\rho L}^{12})^*$ 
and $(g_{\rho L}^{33})^* = g_{\rho L}^{33}$).
The reason why we adopted this texture is as follows:
\begin{itemize}
\item
The size of the real part for $g_{\rho L}^{12}$ actually turns out to be constrained severely 
by the Kaon system measurements such as 
the indirect CP violation $\epsilon_K$, 
and $K_L \to \mu^+ \mu^-$, to be extremely tiny $(\lesssim {\cal O}(10^{-6}))$ 
(for instance, see Ref.~\cite{Buras:2015jaq}). 
In contrast, however, its imaginary part can be moderately larger, which 
will account for the reported $\epsilon'/\epsilon$ anomaly.   
Hence we will take it to be pure imaginary: 
\begin{align} 
{\rm Re} [g_{\rho L}^{12}]=0\,, \qquad 
{\rm Im} [g_{\rho L}^{12}]  \to + g_{\rho L}^{12}
\qquad 
{\rm with} \quad  g_{\rho L}^{12} \in {\bf R}
\,
\,, 
\end{align} 
by which the new physics contributions will be vanishing for 
the $\epsilon_K$ and ${\rm Br}[K_L \to \mu^+ \mu^-]$. 

\item
The base transformation among the {gauge- and flavor-eigenstates}
 can be made by rotating fields as 
(under {the} assumption that neutrinos are massless) 
\begin{align}
{(u_L)^i   = U^{iI} (u'_L)^I,\quad
(d_L)^i   = D^{iI} (d'_L)^I,\quad
(e_L)^i   = L^{iI} (e'_L)^I,\quad
(\nu_L)^i = L^{iI} (\nu'_L)^I,}
	\label{eq:definition_masseigenstates}
\end{align}
where $U$, $D$ and $L$ stand for $3 \times 3$ 
unitary matrices and the spinors with the prime symbol 
denote the fermions in the mass basis, which are specified by    
the capital Latin indices $I$ and $J$. 
The {Cabibbo--Kobayashi--Maskawa (CKM)} matrix is then given by $V_{\text{CKM}} \equiv U^\dag D$
(where corrections to $V_{\rm CKM}$ are ${\cal O}(m_W^2/m_\rho^2)$, being negligible
as long as the CFVs are on the order of TeV). 
As in the literature~\cite{Bhattacharya:2016mcc}, 
we take the mixing structures of $D$ and $L$ as 
\begin{align}
D =
\begin{pmatrix}
{1} & 0 & 0 \\
0 &  \cos{\theta_D} & \sin{\theta_D} \\
0 & -\sin{\theta_D} & \cos{\theta_D} 
\end{pmatrix}, 
\quad\quad
L =
\begin{pmatrix}
{1} & 0 & 0 \\
0 &  \cos{\theta_L} & \sin{\theta_L} \\
0 & -\sin{\theta_L} & \cos{\theta_L}
\end{pmatrix},
\label{LDrotation}
\end{align}
where we {recall} that 
the up-quark mixing matrix is automatically determined through $V_{\text{CKM}} = U^\dag D$.
\end{itemize}

\section{Addressing $B$ anomaly}

When we try to address the anomalies associated with the currents
$b \to s \mu^{+} \mu^{-}$ and $b \to c \tau^- \overline{\nu}$,
we should take account of related constraints from
$B^0_s \,(=s\overline{b}) \leftrightarrow \overline{B^0_s} \,(=b\overline{s})$, $\tau^- \to \mu^- \mu^+ \mu^-$,
$b \to s \nu \overline{\nu}$, $\tau \to \mu s \overline{s}$, and also others.
In our setup, $B^0_s  \leftrightarrow \overline{B^0_s}$, $\tau^- \to \mu^- \mu^+ \mu^-$
generate stringent bounds on the parameters, while constraints from the others are weaker.

\begin{figure}[ht]
\begin{center}
\includegraphics[width=0.5\columnwidth]{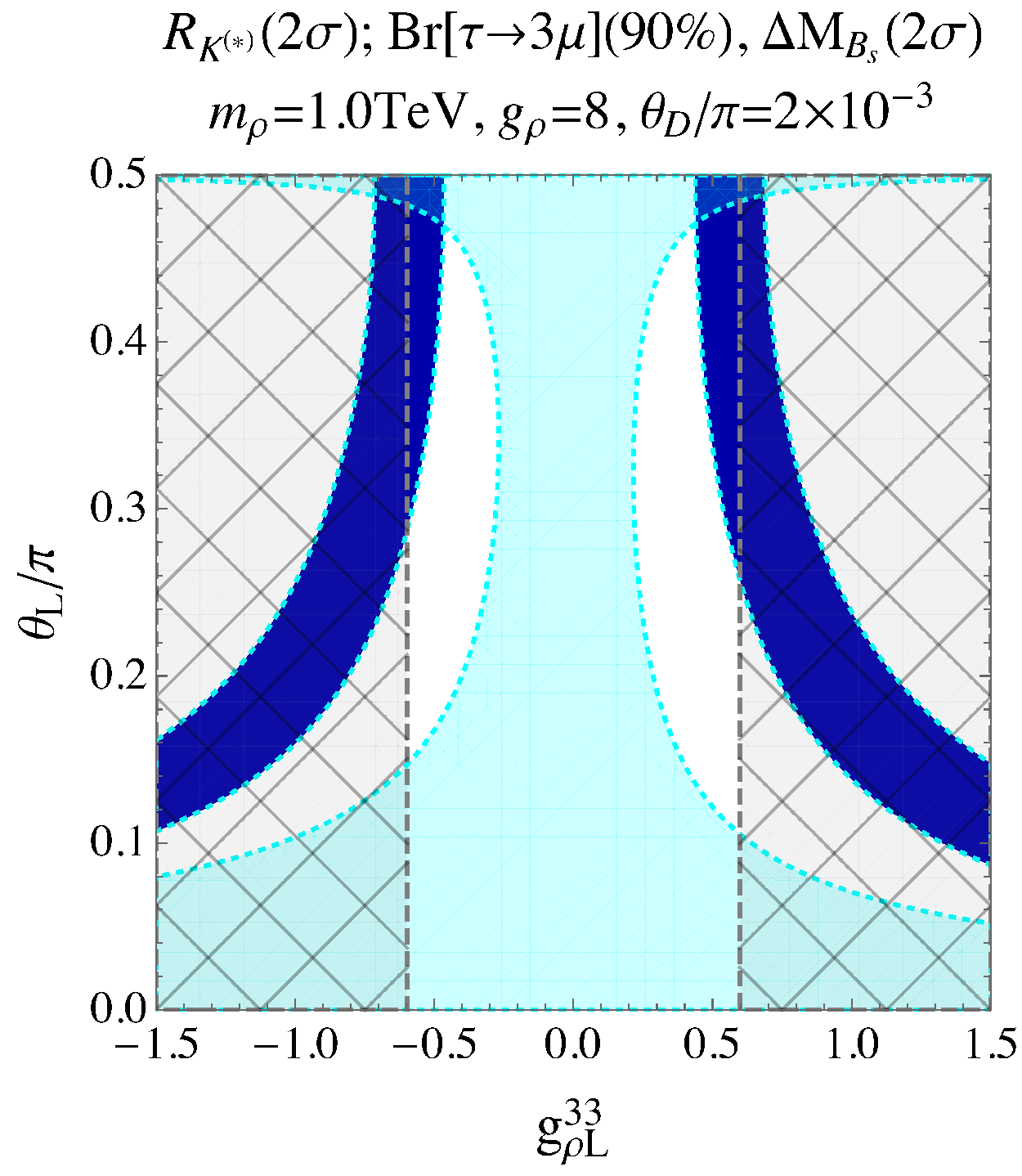}
\caption{
The region plot in the plane 
{$(g_{\rho L}^{33}, \theta_L)$} with {$\theta_D/\pi 
= 2 \times 10^{-3}$} fixed for $m_\rho = 1$ TeV and $g_\rho = 8$. 
The current $R_{K^{(*)}}$ anomaly can be explained in the thick-blue region
at the 2$\sigma$ level, while the cyan-shaded area represents the consistent region
with the current 90\% C.L. upper limit of $\text{Br} [\tau^- \to \mu^- \mu^+ \mu^-]$ (based on the experiment~\cite{Hayasaka:2010np}).
The gray-hatched region is out of the {2$\sigma$-favored} area for $\Delta M_{B_s}$ (based on the result~\cite{Amhis:2014hma}).
We adopted the target magnitude of relevant Wilson coefficients derived through their global fit in~\cite{Capdevila:2017bsm}.
}
\label{B-tau-sys-cons-SU(8)inv-grhoL-rev}
\end{center}
\end{figure}

As a reasonable benchmark, we selected $m_\rho$ and $g_\rho$ as $1\,\text{TeV}$ and eight.
A preferred situation is focused and summarized in Fig.~\ref{B-tau-sys-cons-SU(8)inv-grhoL-rev},
where the down-quark mixing angle $\theta_D$ is chosen as a minuscule digit $(2 \pi \times 10^{-3})$
to circumvent the tight bound from the $B^0_s \leftrightarrow \overline{B^0_s}$ mixing
(the FLAG17 result, $f_{B_s}\sqrt{\hat B_{B_s}}=(274 \pm 8)$ MeV~\cite{Aoki:2016frl}, being adopted
as pointed out in~\cite{DiLuzio:2017fdq}).
Here we have taken into account the NLO QCD operator running effects.
When we focus on the region $(|g_{\rho L}^{33}|, \theta_L) \simeq (0.6, \pi/2)$,
the $R_{K^{(*)}}$ anomaly can be addressed near the best-fit point.

Also, a fascinating property is found in the present scenario in the $R_{D^{(*)}}$ variables.
The setup contains vector-leptoquark-like and $W'$-like vectors, where they can contribute
to $R_{D^{(*)}}$ at the leading order.
However, the underlying $SU(8)$ global symmetry reduces the contribution in total,
which is vanishing in the degenerated limit of CFVs.
This gives us a strong prediction on $R_{D^{(*)}}$ that they are very close to those of SM,
and it will be checked in future by experiments.

\section{Addressing $K$ anomaly}

We move on to the anomaly and associated constraints in the $K$ system. 
Looking at the flavor texture in Eq.(\ref{grhoL}), 
we find that 
the contributions of CFVs to the $s$-$d$ transition observables, 
$\epsilon'/\epsilon$, $K \to \pi \nu {\bar \nu}$ 
and $K^0$-$\bar{K}^0$ mixing
($\Delta M_K$) are possibly generated.
Also, we ought to focus on the $c$-$u$ transition, where observables in the $D^0$-$\bar{D}^0$ mixing
provide us additional constraints on the scenario.

First, we focus on the correlation between $\epsilon'/\epsilon$ and $K^0$-$\bar{K}^0$ mixing ($\Delta M_K$).
For the $\Delta M_K$, 
as was prescribed in Ref.~\cite{Endo:2016tnu},  
we may derive the limit simply by allowing 
the new physics effect to come within the 2$\sigma$ uncertainty 
of the current measurement 
($ \Delta M_K^{\rm exp} = (3.484 \pm 0.006) \times 10^{-15}$ 
GeV~\cite{Tanabashi:2018oca}).
The concrete digit is $|\Delta M_K^{\rm NP} | < 3.496 \times 10^{-15}\,{\rm GeV}$ (NP: new physics).

\begin{figure}[ht]
\begin{center}
\includegraphics[width=0.5\columnwidth]{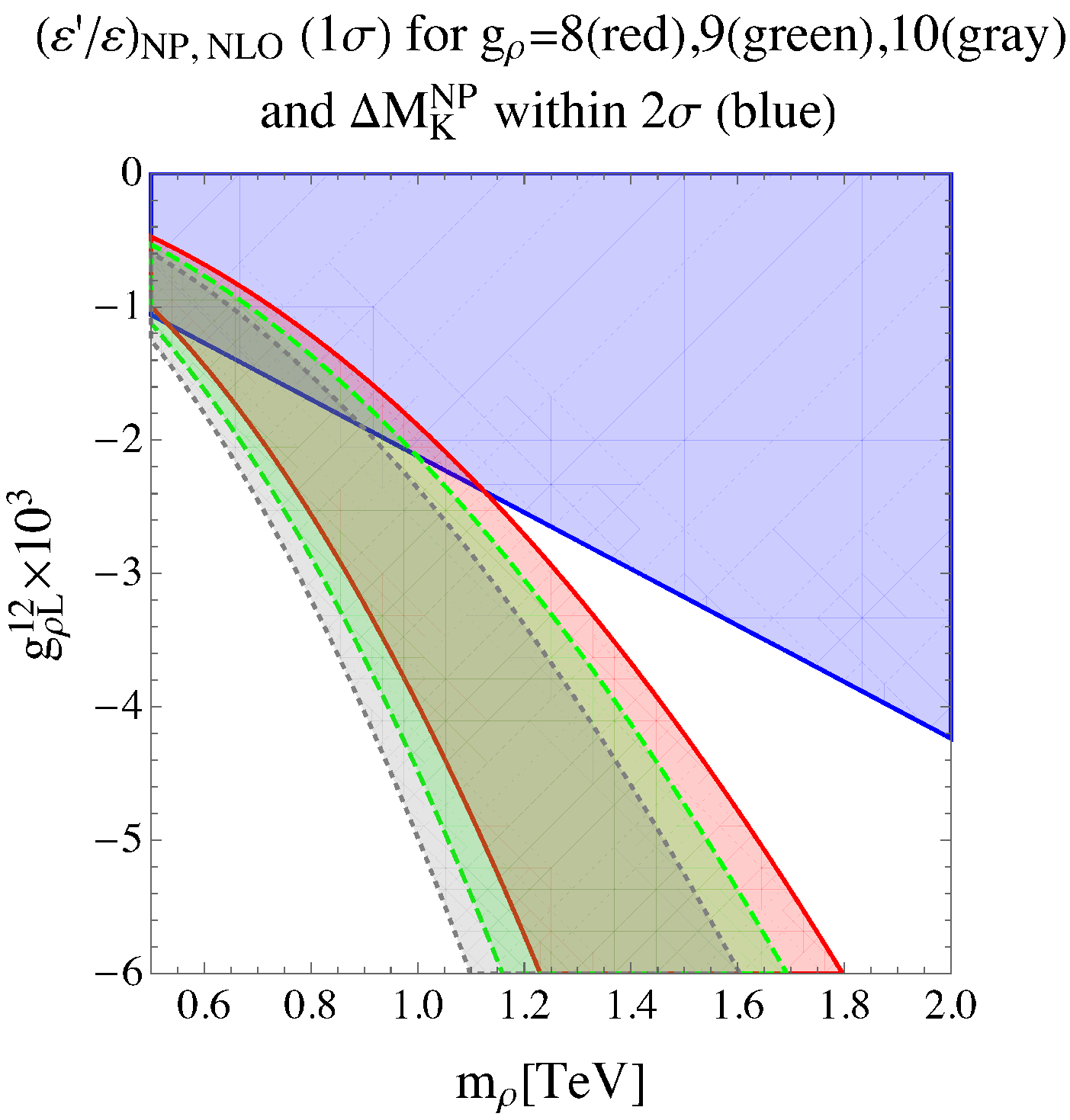}
\caption{
The $\Delta M_K^{\rm NP}$ and $(\epsilon'/\epsilon)_{\rm NP}$ (NP: new physics) constraints 
on the $(m_\rho, g_{\rho L}^{12})$ plane {for the three benchmark values of $g_\rho$}. 
}
\label{epsp-deltaMK-constraint}
\end{center}
\end{figure}

The CFV contributions to the direct CP violation in the $K \to \pi \pi$ processes
are evaluated {at the NLO perturbation in QCD and QED coupling expansions} as \cite{Kitahara:2016nld}
\begin{align} 
 \left(\frac{\epsilon'}{\epsilon}  \right)^{\rm CFVs} 
& =
\frac{\omega_{+}}{\sqrt{2} \left| \epsilon_{K}^{\textrm{exp}}\right|
  \textrm{Re} A_0^{\textrm{exp}}} 
  \langle \vec{Q}_{\epsilon'} ( \mu)^T\rangle 
  \hat{U} \left(\mu, m_{\rho} \right)
 \textrm{Im} \left[
  \vec{C}(m_{\rho}) \right],
\label{epp}
\end{align} 
where $\vec{C}(m_{\rho}) =
(C_1(m_{\rho}),C_2(m_{\rho}),C_3(m_{\rho}), \cdots  , C_{10}(m_{\rho}))^T$
shows the vector from of relevant Wilson coefficients
(see \cite{Matsuzaki:2018jui} for details),
${\textrm{Re}}A_0^{\textrm{exp}} = (3.3201 \pm 0.0018) \times 10^{-7} {\textrm{GeV}}$ \cite{Blum:2015ywa}, 
and
 $\omega_+|_{\rm SM} \equiv a \, {{\rm Re}A_2|_{\rm SM}}/{\rm Re}A_0|_{\rm SM} =
 4.53 
 \times 10^{-2}$ \cite{Cirigliano:2003gt,Buras:2015yba}.
The coefficients $\langle \vec{Q}_{\epsilon'} ( \mu)^T \rangle \hat{U} \left(\mu, m_{\rho} \right)$, which 
denote the evolution of the hadronic matrix elements from the scale $\mu$ to 
the NP scale $m_\rho$,  
are given in Ref. \cite{Kitahara:2016nld}, {where $\langle \vec{Q}_{\epsilon'} ( \mu)^T\rangle$ is defined as}
\begin{align}
{\langle \vec{Q}_{\epsilon'} ( \mu)^T\rangle \equiv
	\frac{1}{\omega_+} \langle \vec{Q}( \mu)^T\rangle_2 - \langle \vec{Q}( \mu)^T\rangle_0 (1 - \hat{\Omega}_{\text{eff}}).}
\end{align}
{The vector forms $\langle \vec{Q}( \mu)^T\rangle_{I}\,(I=0,2)$ are defined from $\langle Q_j({\mu})  \rangle_I$ like $\vec{C}(m_{\rho})$
(where {the concrete information on $\langle \vec{Q}( \mu)^T\rangle_{I}$ and 
$\hat{U} \left(\mu, m_{\rho}\right)$ are available in~\cite{Kitahara:2016nld}}).
The factors for the isospin breaking correction are described in the matrix form,}
\begin{align}
{(1 - \hat{\Omega}_{\text{eff}})_{ij}
	=
\begin{cases}
0.852 & (i = j = 1-6), \\
0.983 & (i = j = 7-10), \\
0        & (i \not= j).
\end{cases}}
\end{align}
{Here the scale $\mu$ is set to be $1.3\,\text{GeV}$.}
In the LO analysis where $C_5(m_{\rho}), C_6(m_{\rho})$ and $C_7(m_{\rho})$
bring main effects on $C_6(m_c)$ and $C_8(m_c)$,  
we found that the contributions from QCD penguin $Q_6$ dominates in the $\epsilon'/\epsilon$, 
and the EW penguin $Q_8$ term yields about $60 \%$ contribution of them. 
Fig.~\ref{epsp-deltaMK-constraint} tells us that there is an upper bound on $m_\rho$ for each choice of $g_\rho$,
e.g., $m_\rho \lesssim 1.4\,\text{TeV}$ for $g_\rho = 8$.
For addressing the $(\epsilon'/\epsilon)$ anomaly, we need nonzero $\mathcal{V}$-$\rho$ mixing to generate
relevant connections to current interactions, where we remind that $(g_{s,W,Y}/g_\rho)$ determines the size of the mixing.
This property gives us a great hint for pursuing signals of this scenario in experiments.

\begin{figure}[ht]
\begin{center}
\includegraphics[width=0.45\columnwidth]{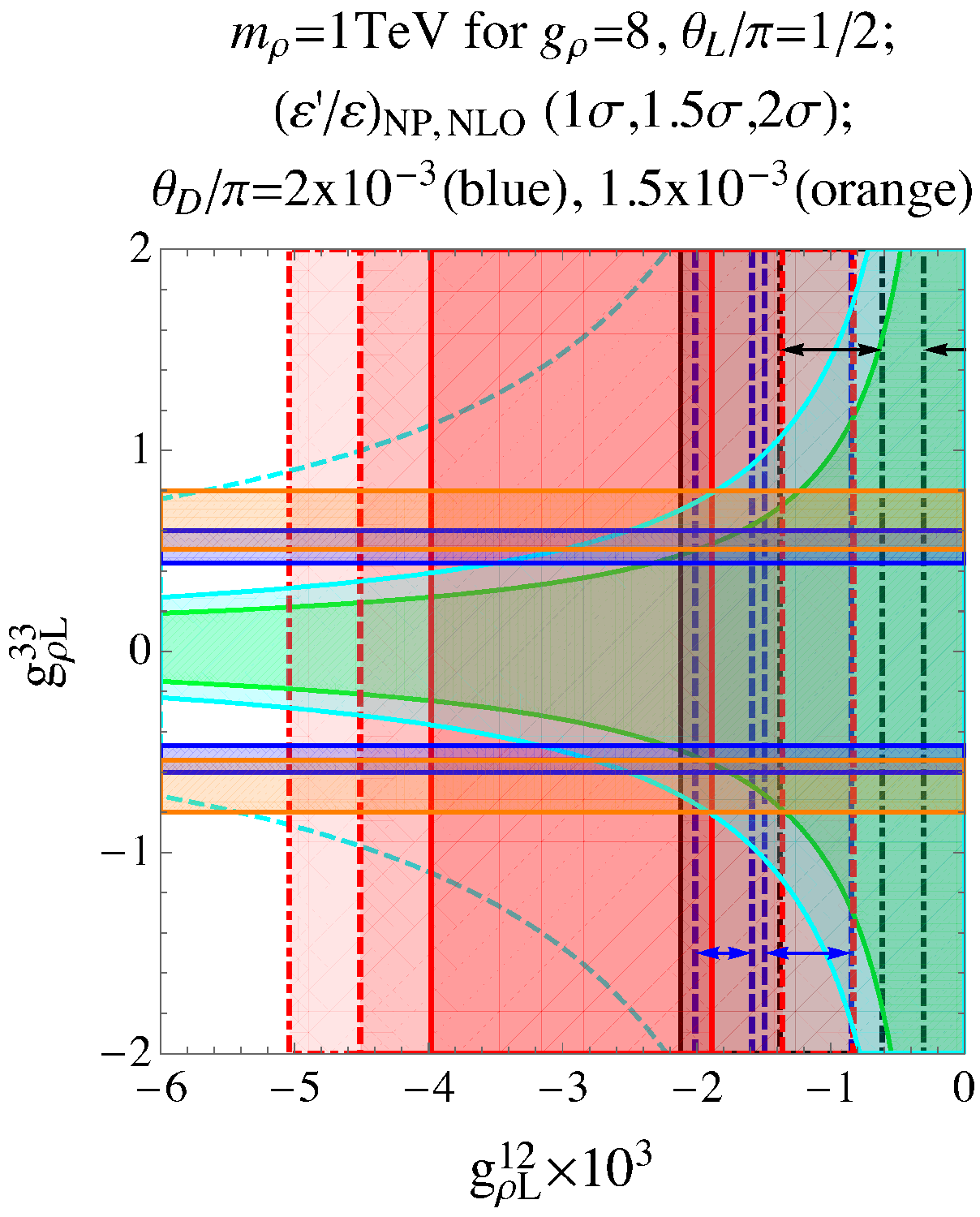}
\includegraphics[width=0.45\columnwidth]{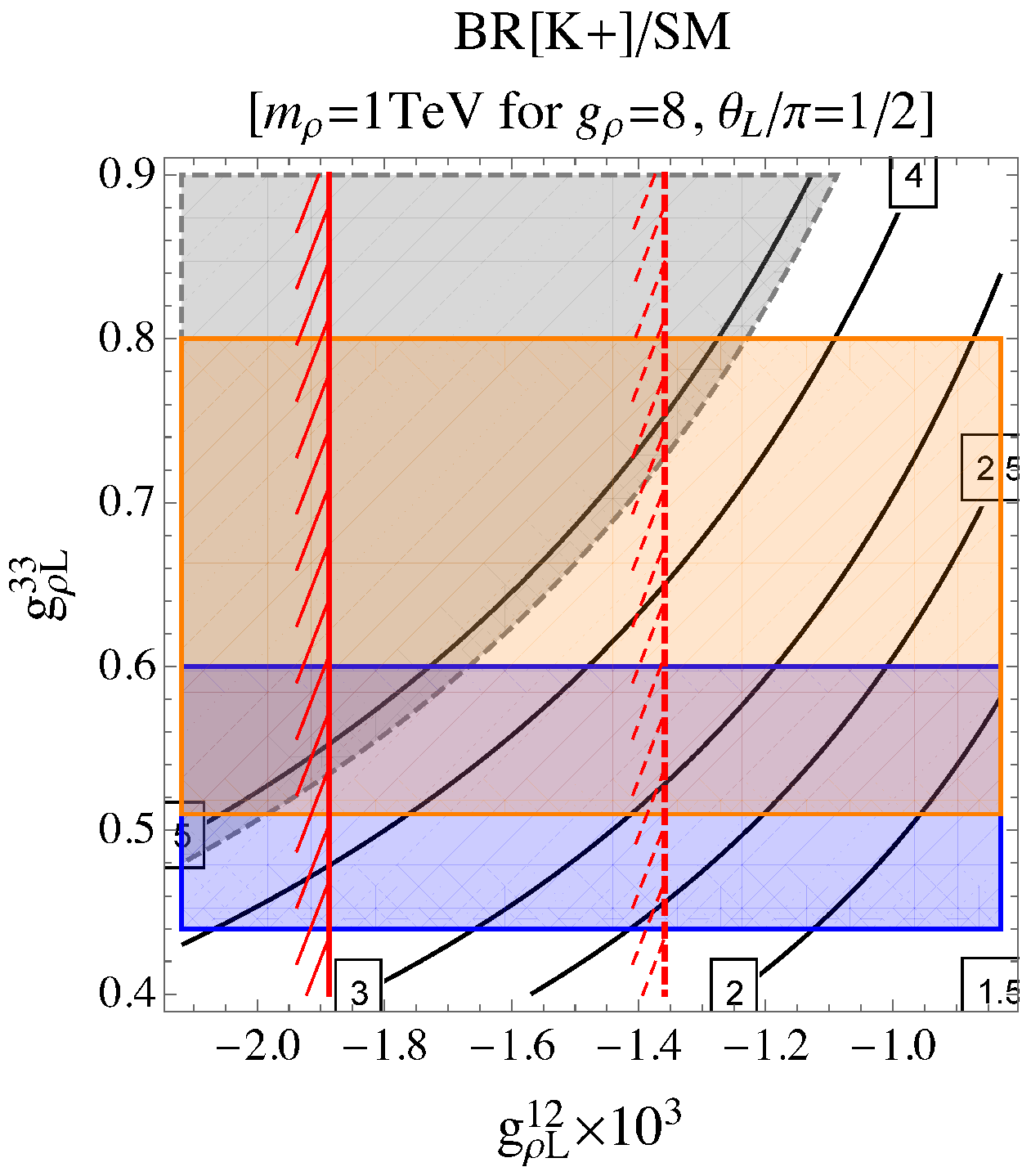} 
\caption{
\textbf{Left}:
the combined constraint plot on 
{$(g_{\rho L}^{12}, g_{\rho L}^{33})$} for $m_\rho = 1$ TeV, $g_\rho = 8$, {$\theta_L/\pi = 1/2$ and $\theta_D/\pi = 2 \times 10^{-3}$
(horizontal band in blue) or $1.5 \times 10^{-3}$ (in orange)}, 
where the shaded regions are allowed. 
The red and pale-black vertical domains respectively 
correspond to the allowed regions set by {the $1\sigma$ (surrounded by solid line boundaries), $1.5\sigma$ (by dashed ones), $2\sigma$ 
(by dot-dashed ones) ranges} for $(\epsilon'/\epsilon)_{\rm NP}$,   
and the $2\sigma$ range for $\Delta M_K$. 
The $2\sigma$-allowed range for Br[$K^+ \to \pi^+ \nu \bar{\nu}$] (based on the experimental result~\cite{Artamonov:2008qb})
and the 90\% C.L. upper bound for Br[$K_L \to \pi^0 \nu \bar{\nu}$] (based on the experimental results~\cite{Ahn:2009gb,Ahn:2018mvc})
have been reflected in domains wrapped by green and cyan regions, respectively.  
The upper bound on Br[$K_L \to \pi^0 \nu \bar{\nu}$] was updated by the KOTO experiment at ICHEP in July 2018~\cite{Ahn:2018mvc}. 
In the figure we have also shown 
the previous boundary based on the previous bound by the dashed cyan curves.
The regions surrounded by horizontal lines 
[in blue (for $\theta_D/\pi = 2 \times 10^{-3}$) or orange (for $\theta_D/\pi = 1.5 \times 10^{-3}$)] are allowed by the $B-\tau$ system constraint 
in Fig.~\ref{B-tau-sys-cons-SU(8)inv-grhoL-rev}, 
in which 
the lower bounds on the magnitude of $g_{\rho L}^{33}$ 
come from the requirement to account for the 
$R_{K^{(*)}}$ anomaly within the $2\sigma$ level, while the upper ones originate from circumventing the bound from 
$\Delta M_{B_s}$ at the $2\sigma$ level, respectively.
The vertical domains identified by the blue and black horizontal arrows correspond
to the $95\%$ C.L. intervals of the ${D}^0$-$\bar{D}^0$ mixing
when $x^\text{SM} = +1\%$ and $+0.1\%$, respectively (see~\cite{Matsuzaki:2018jui} for details).
\textbf{Right}:
the magnified plot of the left panel focused on the region where both of the $R_{K^{(*)}}$ and $\epsilon'/\epsilon$ anomalies
can be addressed with positive $g_{\rho L}^{33}$.
The gray-shaded region is already excluded by Br[$K^+ \to \pi^+ \nu \bar{\nu}$].
The black contours describes the excess as Br[$K^+ \to \pi^+ \nu \bar{\nu}$]${}^{\text{CFVs}}$$/$Br[$K^+ \to \pi^+ \nu \bar{\nu}$]${}^{\text{SM}}$.
} 
\label{K-sys-cons-SU(8)inv-grhoL-rev}
\end{center}
\end{figure}

Second, we go for the summary plot, where the following parameters,
$(m_\rho, g_\rho, \theta_L, \theta_D)$, are fixed as
$(1\,\text{TeV}, 8, \pi/2, \, 2 \pi \times 10^{-3} \text{ or } 1.5 \pi \times 10^{-3})$, respectively
to address the $R_{K^{(*)}}$ anomaly successfully.
In addition to $\Delta M_K$, Br[$K^+ \to \pi^+ \nu \bar{\nu}$], Br[$K_L \to \pi^0 \nu \bar{\nu}$], and
$D^0$-$\bar{D}^0$ mixing provide constraints on the scenario.
We itemize the relevant aspects of Fig.~\ref{K-sys-cons-SU(8)inv-grhoL-rev}.
\begin{figure}[ht]
\begin{center}
\includegraphics[width=0.6\columnwidth]{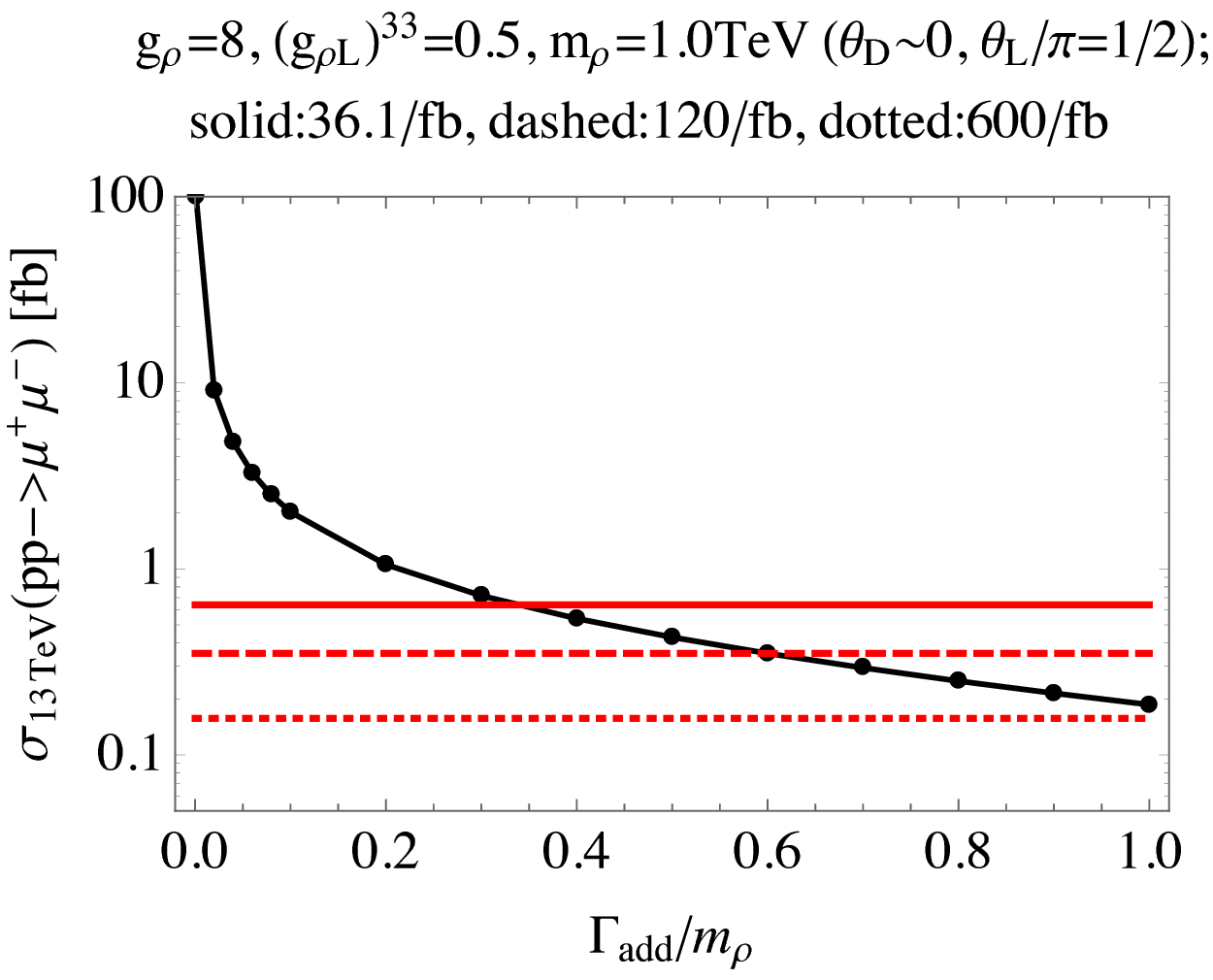}
\caption{
The dimuon resonant production cross section for the target 
CFVs ($\rho_{(1)}^3$ and $\rho_{(1)^\prime}^0$) at LHC with $\sqrt{s}=$13 TeV 
as a function of a possibly added width term (common for two CFVs) 
normalized to the mass $m_\rho$,  
for $m_\rho$= 1 TeV, $g_\rho=8$, {$g_{\rho L}^{33}=0.5$ (and $\theta_D \sim 0$, $\theta_L = \pi/2$)}. 
The horizontal {solid, dashed and dotted} lines (in red)
respectively correspond to the current 95\% C.L upper limit 
placed by the ATLAS group with the integrated luminosity 
${\cal L}=36.1 \,{\rm fb}^{-1}$~\cite{ATLAS:2017wce}, 
and the expected upper bounds at ${\cal L}=120 \,{\rm fb}^{-1}$ 
and {${\cal L}=600\,{\rm fb}^{-1}$}   
estimated just by {simply} scaling the luminosity.
The LHC cross section has been computed by implementing 
the {\tt CTEQ6L1} {parton distribution function~(PDF)}~\cite{Pumplin:2002vw} 
in {\tt Mathematica} with the help of a PDF parser package, 
{\tt ManeParse\_2.0}~\cite{Clark:2016jgm}, and setting 
{$\tau_0\equiv 4 m^2_{\rm threshold}/s= 10^{-6}$} 
as the minimal value of the Bjorken $x$ in the {\tt CTEQ6L1} PDF set, 
where the PDF scale is set to $m_\rho$.
The {\tt CUBA} package~\cite{Hahn:2004fe} {has been}
utilized for numerical integrations.
}
\label{with-Gamma-1TeV-rev}
\end{center}
\end{figure}
\begin{itemize}
\item
Since the flavor of the neutrinos is not identified in experiments,
the constraints from $K^+ \to \pi^+ \nu \bar{\nu}$ and $K_L \to \pi^0 \nu \bar{\nu}$
hold correlations between $g_{\rho L}^{12}$ and $g_{\rho L}^{33}$.
\item
Addressing $R_{K^{(*)}}$ and $\epsilon'/\epsilon$ within $1\sigma$ C.L.s looks possible, but it is realizable
very in a limited parameter space.
If we relax it to $2\sigma$ C.L.s, still we find sufficient regions of parameters.
\item
A part of the region where $\epsilon'/\epsilon$ can be addressed
seems to be excluded by the constraint from the $D^0$-$\bar{D}^0$ mixing.
Nevertheless, as shown in the left panel of Fig.~\ref{K-sys-cons-SU(8)inv-grhoL-rev}, the $95\%$ C.L.
interval highly depends on the choice of the uncertain input $x^\text{SM}$ (see~\cite{Matsuzaki:2018jui} for details).
Taking account of the uncertainty for $x^\text{SM}$, we can conclude that no definite bound is put on the red vertical domains
(for $\epsilon'/\epsilon$)
in Fig.~\ref{K-sys-cons-SU(8)inv-grhoL-rev}.
\item
According to the literature~\cite{NA62:2017rwk}, 
by the end of 2018 the NA62 experiment will measure  
the $K^+ \to \pi^+ \nu \bar{\nu}$ with about 10\% accuracy 
of the SM prediction.
The KOTO experiment also plans to report new results on the 
data analysis on the $K_L \to \pi^0 \nu \bar{\nu}$ in the near future, 
to be expected to reach the level of $< 10^{-9}$ for the branching ratio, 
corresponding to 2015 - 2018 data taking~\cite{KOTO2018}.
In the near future, such updated bounds will restrict the relevant parameter regions sizably.
\end{itemize}

\section{Status of LHC dilepton search}

Generally, vector particles coupled with a pair of muons are significantly constrained by direct dimuon resonance searches at the LHC.
In our scenario, flavor universal couplings to up- and down-quarks dominate production cross section even though
the suppression factor $(g_{s}/g_\rho)$ works.
As shown in Fig.~\ref{with-Gamma-1TeV-rev}, the additional widths of the CFVs contributing the production are required to be large as
$\Gamma_\text{add}/m_\rho \sim 30\%$.
Such an additional width would be present 
when the CFV can dominantly couple to a 
hidden dark sector including a dark matter candidate, 
or a pionic sector realized 
as in a hidden QCD with a setup similar to the present CFV 
content~\cite{Matsuzaki:2017bpp}.

\section{Summary}

In the present setup of chiral-flavorful vectors~(CFVs), not only the (assumed) flavor changing interactions,
but also the flavor universal interactions through the mixing of the CFVs and the gauge bosons in the standard model~(SM)
due to manifestly realized SM gauge invariance, play important roles.
Under the presence of the latter interactions, we can successfully address the anomaly in $\epsilon'/\epsilon$ in the Kaon sector,
in addition to the deviations in observables associated with $b \to s \mu^+ \mu^-$, including $R_{K^{(*)}}$.
Almost vanishing contributions to the variables $R_{D^{(*)}}$ are realized due to remnant of the $SU(8)$ global symmetry,
which is a strong prediction of our scenario even though the presently reported excesses in $R_{D^{(*)}}$ cannot be explained.
Limited (quite wide) parameter space still remains for a simultaneous explanation of $R_{K^{(*)}}$ and $\epsilon'/\epsilon$
within $1\sigma$ ($2\sigma$) C.L. individually.
In addition to the LHC direct searches, also the NA62 and KOTO experiments will explore the parameter space for the explanations
through measuring Br[$K^+ \to \pi^+ \nu \bar{\nu}$] and Br[$K_L \to \pi^0 \nu \bar{\nu}$], respectively.

\bigskip

\noindent
{\bf Note added}:
In the latter half of March 2019 during the Rencontres de Moriond 2019,
the Belle and LHCb experimental groups have published their latest results on the $B$ anomalies, which include
\begin{align}
R_{K^{*}}^{[0.045, 1.1]}  &= 0.52^{+0.36}_{-0.26} \pm 0.05,  & &\text{(Belle)~\cite{Belle-RK-Moriond2019}}, \\
R_{K^{*}}^{[1.1, 6]}         &= 0.96^{+0.45}_{-0.29} \pm 0.11,  & &\text{(Belle)~\cite{Belle-RK-Moriond2019}}, \\
R_{K}^{[1.1, 6]}              &= 0.846^{+0.060+0.016}_{-0.054-0.014},  & &\text{(LHCb)~\cite{LHCb-RK-Moriond2019}}, \\
R_{D}      &= 0.307 \pm 0.037 \pm 0.016,  & &\text{(Belle)~\cite{Belle-RD-Moriond2019}}, \\
R_{D{*}}  &= 0.283 \pm 0.018 \pm 0.014,  & &\text{(Belle)~\cite{Belle-RD-Moriond2019}}.
\end{align}
The LHCb group announced that their updated result on $R_K$ by use of $2011$, $2012$, $2015$ and $2016$ data
becomes $\sim 2.5\,\sigma$ from the SM (previously $\sim 2.6\,\sigma$ using $2011$ and $2012$ data), while
the Belle group claimed that the $R_{D}$-$R_{D^{*}}$ experimental world average of the deviation is decreased to
$3.1\,\sigma$ from $3.8\,\sigma$.
These results implies that both of the $R_{K^{(*)}}$ and $R_{D^{(*)}}$ anomalies may become `less significant'.
Nevertheless, sizable deviations can remains, and then the discussion on this manuscript is basically still valid.

\section*{Acknowledgements} 

The presenter (K.N.) would like to thank the organizers for kind invitation and hospitalities during the workshop.

\end{document}